\documentclass[lettersize,journal]{IEEEtran}

\usepackage{cite}
\usepackage{amsmath,amssymb,amsfonts}
\usepackage{algorithmic}
\usepackage{graphicx}
\usepackage{textcomp}
\usepackage{booktabs,multirow}
\usepackage{tabularx}
\usepackage[table,dvipsnames,svgnames,x11names]{xcolor}
\usepackage{colortbl}
\usepackage{hhline}
\usepackage{arydshln}
\usepackage{relsize}
\usepackage{float}
\usepackage{url}
\usepackage{threeparttable}
\usepackage{pifont}
\usepackage{marvosym}
\usepackage{ragged2e}
\usepackage{multirow}
\usepackage{vcell}
\usepackage{courier}
\usepackage{booktabs}
\usepackage{soul}
\usepackage{comment}
\usepackage[hidelinks,colorlinks=true,linkcolor=Emerald,citecolor=cyan,urlcolor=black]{hyperref}
\usepackage{soul}

\usepackage{balance}

\usepackage{pgfplots}
\usepgfplotslibrary{groupplots}
\pgfplotsset{compat=1.18}
\usepgfplotslibrary{statistics}

\usepackage{tikz,xcolor}

\definecolor{lime}{HTML}{A6CE39}
\DeclareRobustCommand{\orcidicon}{%
	\begin{tikzpicture}
	\draw[lime, fill=lime] (0,0) 
	circle [radius=0.14] 
	node[white] {{\fontfamily{qag}\selectfont \tiny ID}};
	\draw[white, fill=white] (-0.0625,0.095) 
	circle [radius=0.007];
	\end{tikzpicture}
	\hspace{-2mm}
}

\foreach \x in {A, ..., Z}{%
	\expandafter\xdef\csname orcid\x\endcsname{\noexpand\href{https://orcid.org/\csname orcidauthor\x\endcsname}{\noexpand\orcidicon}}
}

\newcommand{\ignore}[1]{}

\newcommand{\Design}{\textbf{\texttt{FALCON}}}

\usepackage{setspace}
\begin{document}

\title{%MICASA: Fault-Tolerant MTJ-Based Stochastic In-Memory Computing Architecture for Edge-AI
%\huge FALCON: Fault-Tolerant MTJ-Based Stochastic In-Memory Computing Architecture for Edge-AI
 FALCON: Fault-Tolerant MTJ-Based In-Memory Stochastic Architecture for Reliability-Critical\\ Edge AI Applications
%\vspace{-0.25em}
}

\author{
\IEEEauthorblockN{Farzad Razi\IEEEauthorrefmark{1}\orcidA,~\IEEEmembership{Member,~IEEE}, Mehran Moghadam\IEEEauthorrefmark{1}\orcidB,~\IEEEmembership{Graduate Member,~IEEE}}, Sercan Aygun\orcidC,~\IEEEmembership{Senior Member,~IEEE}, M. Hassan Najafi\orcidD,~\IEEEmembership{Senior Member,~IEEE}, and Marc Riedel\orcidE,~\IEEEmembership{Senior Member,~IEEE}%
\vspace{-1.5em}
\thanks{%\scriptsize
%This work is supported by NSF grants 2019511, 2339701, and 2609436, NASA grant 80NSSC25C0335, Vernon $\&$ Ruby Langlinais Non-Endowed Research Fund, Lockheed Martin Corporation Endowed Professorship Fund, and gifts from NVIDIA and Google.
Farzad Razi and Marc Riedel are with the ECE Department, University of Minnesota, Minneapolis, MN, USA (e-mail:\{frazi,mriedel\}@umn.edu). Mehran Moghadam and M. Hassan Najafi are with the ECSE Department, Case Western Reserve University, Cleveland, OH, USA (email:\{moghadam,najafi\}@case.edu). Sercan Aygun is with the School of Computing and Informatics, University of Louisiana at Lafayette, Lafayette, LA, USA (e-mail:sercan.aygun@louisiana.edu). (\IEEEauthorrefmark{1}Equal contribution.)}%
%\thanks{\IEEEauthorrefmark{1}The authors contributed equally to this work.}
%\vspace{-0.5em}
}

%\markboth{IEEE TRANSACTIONS ON VERY LARGE SCALE INTEGRATION (VLSI) SYSTEMS, Vol.~XX, No.~X, XXXX~2026}%

\maketitle

\begin{abstract}
As modern data-centric applications such as neural inference and sensor-edge analytics expand, they increasingly encounter the von Neumann memory wall, suffering from excessive data movement overhead and stringent energy constraints. In-Memory Computing (IMC) utilizing emerging non-volatile technologies, such as Magnetic Tunnel Junctions (MTJs), promises to mitigate these bottlenecks. However, conventional binary radix-based IMC architectures suffer from excessive vulnerability to process-induced variations, restricted operating margins, and thermal noise. To bridge the gap between \textit{energy efficiency} and \textit{computational reliability}, this work proposes \Design{}, a fault-tolerant, MTJ-based in-memory arithmetic architecture integrated with \textit{Stochastic Computing (SC)}. By encoding numerical values into %deterministic %randomly generated 
uniform bit-streams, %\red{[we state in the text that we generate BS deterministically, this could be misleading]}, 
SC naturally absorbs localized soft errors and enables the execution of an essential %comprehensive \red{[or essential?]} 
suite of arithmetic operations using highly compact logic primitives directly within the memory arrays. \Design{} integrates a \textit{deterministic} bit mapping mechanism with reconfigurable logic-in-memory (LIM) structures, eliminating the need to transfer data to external processors or %rely on 
area- and power-hungry random number generators. Experimental results using 14 $nm$ FinFET technology validate the correct functionality of \Design{} %framework 
even under aggressive voltage scaling, severe process variation, and noise injection levels up to 30\%, %and noise resilience up to 30\%, 
making it a robust framework for reliability-critical edge AI applications. We investigate the proper functionality of \Design{} on morphological closing as a realistic noise-tolerant image processing case study.
\end{abstract}

\begin{IEEEkeywords}
Fault tolerance, in-memory computing, %magnetic tunnel junction, 
morphological closing, %reliability, 
stochastic computing.
%\vspace{-1em}
\end{IEEEkeywords}

\section{Introduction and Motivation}
%\vspace{-0.1em}

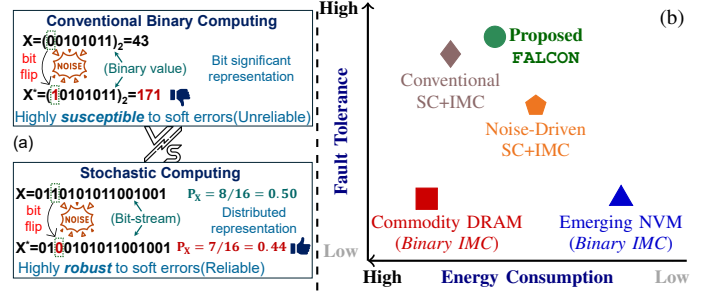
\begin{figure}[!t] % The asterisk (*) makes it span both text columns!
\centering

% ==========================================
% LEFT COLUMN: Stacked Figures
% ==========================================
\begin{minipage}[c]{0.45\columnwidth}
    \centering
    % Top Image (Part a)
    \hspace{-1.8em}\includegraphics[width=\columnwidth]{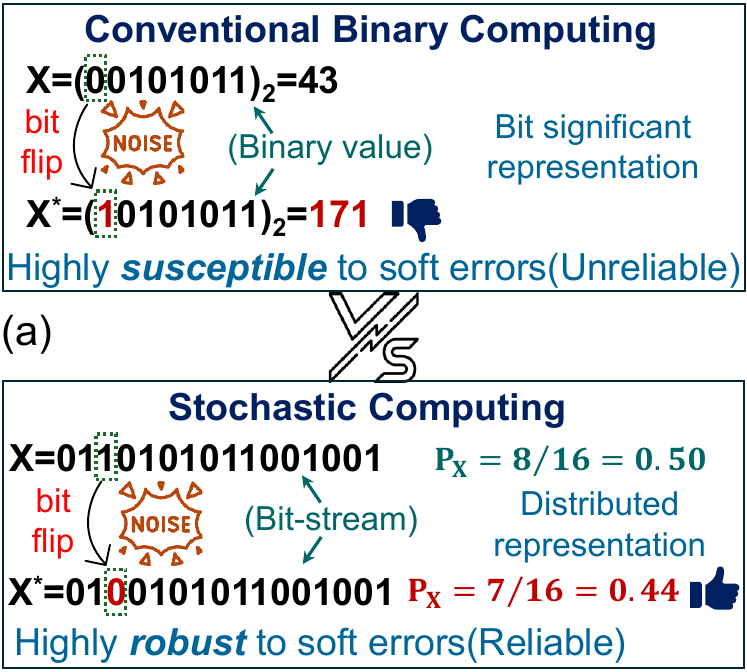}

\end{minipage}
%\hfill % Pushes the two minipages apart to balance them
\hspace{-1.5em}
\begin{tikzpicture}[baseline=(current bounding box.center)]
    % Draws a 7cm tall dashed gray line. 
    % Adjust the '3.5' values if you need the line to be taller or shorter!
    \draw[dashed, thick, black] (0, -1.1) -- (0, 2.6);
\end{tikzpicture}
\hspace{-1em}
% ==========================================
% RIGHT COLUMN: White Space Chart
% ==========================================
\begin{minipage}[c]{0.53\columnwidth}
    \centering
    \begin{tikzpicture}
\begin{axis}[
    width=5.8cm,  % Squeezed physical width 7.2
    height=5.0cm, % Squeezed physical height 5.8
    xmin=1.0, xmax=9.5, % Cropped internal X white space
    ymin=0.5, ymax=8.0, % Stopped exactly at 8.0, padded the bottom
    clip=false,         % Ensures no custom arrows or text get cropped
    x dir=reverse, % Reverses the X-axis so High is on the left, Low is on the right
    axis x line=bottom,
    axis y line=left,
    enlarge x limits=0.05,
    % enlarge y limits=0.05, % REMOVED to align the arrow perfectly with 8.0
    % Customizing the ticks to match the new squeezed limits
    xtick={1.0, 9.5},
    xticklabels={\textcolor{gray!70!white}{Low}, \textcolor{black}{High}}, % 
    ytick={0.8, 7.8},
    yticklabels={\textcolor{gray!70!white}{Low}, \textcolor{black}{High}},
    xlabel={\textcolor{Navy}{\textbf{Energy Consumption}}},%\\ \textcolor{Navy}{\textbf{(Power Overhead)}}},
    xlabel style={align=center, yshift=1.3em},
    ylabel={\textcolor{Navy}{\textbf{Fault Tolerance}}},%\\ \textcolor{Navy}{(Robustness)}},
    ylabel style={align=center, yshift=-1.75em},
    label style={font=\scriptsize\bfseries},
    tick label style={font=\scriptsize\bfseries},
    axis line style={->, thick, black}, % ADDED '->' to draw the axis arrows
    % Removing tick marks but keeping the labels
    major tick length=0pt,
]

% 1. Bottom-Left: Commodity DRAM Binary IMC (High Energy, Low Robustness)
\addplot[mark=square*, mark size=4pt, draw=red!80!black, fill=red!80!black] coordinates {(8.2, 2.3)};
\node[align=center, font=\scriptsize, anchor=north] at (axis cs:7.6, 2.1) {\textcolor{red!80!black}{Commodity DRAM}\\ \textcolor{red!80!black}{(\textit{Binary IMC})}};

% 2. Bottom-Right: Emerging NVM Binary IMC (Low Energy, Low Robustness)
\addplot[mark=triangle*, mark size=5.0pt, draw=blue!80!black, fill=blue!80!black] coordinates {(2.5, 2.3)};
\node[align=center, font=\scriptsize, anchor=north] at (axis cs:2.5, 2.1) {\textcolor{blue!80!black}{Emerging NVM}\\ \textcolor{blue!80!black}{(\textit{Binary IMC})}};

% 3. Top-Left: Conventional SC+IMC with CMOS RNGs (High Energy, High Robustness)
\addplot[mark=diamond*, mark size=5.0pt, draw=RosyBrown4, fill=RosyBrown4] coordinates {(7.5, 6.5)};
\node[align=center, font=\scriptsize, anchor=north] at (axis cs:7.5, 6.2) {\textcolor{RosyBrown4}{Conventional}\\ \textcolor{RosyBrown4}{SC+IMC}};%\\(CMOS RNGs)};

% 4. Top-Middle: Stochastic NVM using Device Noise (Medium Energy, High Robustness)
\addplot[mark=pentagon*, mark size=4.0pt, draw=Chocolate2, fill=Chocolate2] coordinates {(5, 5)};
\node[align=center, font=\scriptsize, anchor=north] at (axis cs:5, 4.8) {\textcolor{Chocolate2}{Noise-Driven}\\ \textcolor{Chocolate2}{SC+IMC}};

% 5. Top-Right (The White Space): FALCON (medium Energy, High Robustness)
\addplot[mark=oplus*, mark size=4.0pt, draw=SeaGreen4, fill=SeaGreen4] coordinates {(6.2, 7)}; % 6.2, 7
\node[align=center, font=\scriptsize\bfseries, text=green!40!black, anchor=north] at (axis cs:4.7, 7.5) % 4.7, 7.5
{Proposed\\ \Design{}};%\\(Deterministic SC)};

\node[align=center, font=\footnotesize, anchor=south] at (axis cs:1, 7) {(b)};

% Drawing an arrow to highlight FALCON's position
%\draw[->, thick, green!40!black] (axis cs:2.6, 7.3) -- (axis cs:1.8, 7.0);
%\node[align=center, font=\scriptsize\itshape, anchor=south] at (axis cs:3.5, 7.1) {High noise-tolerance};%,\\Zero-RNG In-Memory Math};

\end{axis}
\end{tikzpicture}

\end{minipage}

\vspace{-0.5em}
\caption{
(a) Two computing paradigms from \textit{reliability} perspective. (b) State-of-the-art IMC architectures, evaluating their positioning across \textit{energy consumption} and \textit{fault tolerance}.
%Comprehensive overview of the proposed architecture. (Left) State transition diagram and sample outputs for the deterministic bit-stream generation. (Right) White-space chart evaluating the positioning of various IMC designs based on energy efficiency and robustness, illustrating how \Design{} uniquely occupies the critical white space of high fault tolerance and ultra-low energy consumption.
}
\label{SC_binary_comp_white_chart}
\vspace{-1.5em}
\end{figure}

\IEEEPARstart{W}{ith} the continued expansion of modern computing workloads in both scale and complexity, ensuring reliable and error-resilient computation has become a critical design objective alongside performance and energy efficiency.
%As modern computing workloads continue to expand in scale and complexity, ensuring reliable and error-resilient computation has become as critical as improving performance and energy efficiency. 
Applications such as neural inference, sensor-edge analytics, and large-scale data processing frequently operate under strict energy budgets while being exposed to device noise, supply-voltage fluctuations, and fabrication variations. Under these conditions, conventional von Neumann processors suffer from both high data-movement overhead and reduced computational robustness. This motivates paradigms that bring processing %the adoption of computing paradigms in which processing occurs 
closer to data and inherently reduce sensitivity to hardware faults. %where the underlying representations that naturally suppress the impact of hardware unreliability.
In-Memory Computing (IMC) directly addresses the data movement bottleneck by embedding computations %computational capabilities 
inside or near memory arrays~\cite{sun2023survey}. By minimizing transfers between storage and processing units, IMC significantly reduces opportunities for error propagation, timing failures, and transient disturbances. However, when implemented with emerging nanoscale devices such as magnetic tunnel junctions (MTJs), %magnetic tunnel junctions (MTJs), %which are radiation hardened, 
IMC architectures remain vulnerable to \textit{process-induced variations}, \textit{limited operating margins}, and \textit{thermal noise}, all of which can threaten correctness unless the computational model itself is inherently tolerant to imperfections.
Binary IMC architectures are inherently susceptible to soft errors, and can become unreliable even at low soft-error rates~\cite{Alam_IMC_Reliability_GLSVLSI23}. %and they may easily become unreliable in the presence of $>$5\% soft errors~\cite{Alam_IMC_Reliability_GLSVLSI23}.
Stochastic Computing (SC)~\cite{qian2010architecture} offers a fault-tolerant computational model in which numerical values are encoded as uniform %random \red{[deterministic?]} 
bit-streams and arithmetic operations are performed using %extremely 
simple logic primitives (e.g., multiplication via bit-wise \texttt{AND}). This probability-based representation naturally absorbs local errors: a noise-induced bit flip in a stochastic stream can perturb only one least-significant bit and so has minimal impact on the final result~\cite{alaghi2017promise} (see the example in Fig.~\ref{SC_binary_comp_white_chart}(a)). 
%Also the impact of device noise has minimal impact %is naturally averaged out 
%across the entire bit-stream, reducing its significance. 
These characteristics make SC exceptionally well suited to environments where reliability is difficult to guarantee, such as deeply scaled memory-centric architectures. 
%On the other hand, MTJ devices 
The pairing of SC and MTJs is particularly advantageous: by exploiting the intrinsic write/read behavior of MTJs, bit-streams can be generated natively in memory, eliminating a major bottleneck in the cost-efficiency of conventional SC systems, namely the overhead of dedicated bit-stream generation circuitry.
%enable native in-memory bit-stream generation by exploiting their intrinsic write/read behavior,  
%Furthermore, the intrinsic write/read behavior of MTJs enables compact in-memory bit-stream generation, 
%eliminating the need for costly stochastic bit-stream generation circuitry. %traditionally high hardware cost of random number generators in off-memory SC designs.

\begin{figure*}[!t]
    \centering
        \includegraphics[width=\linewidth]{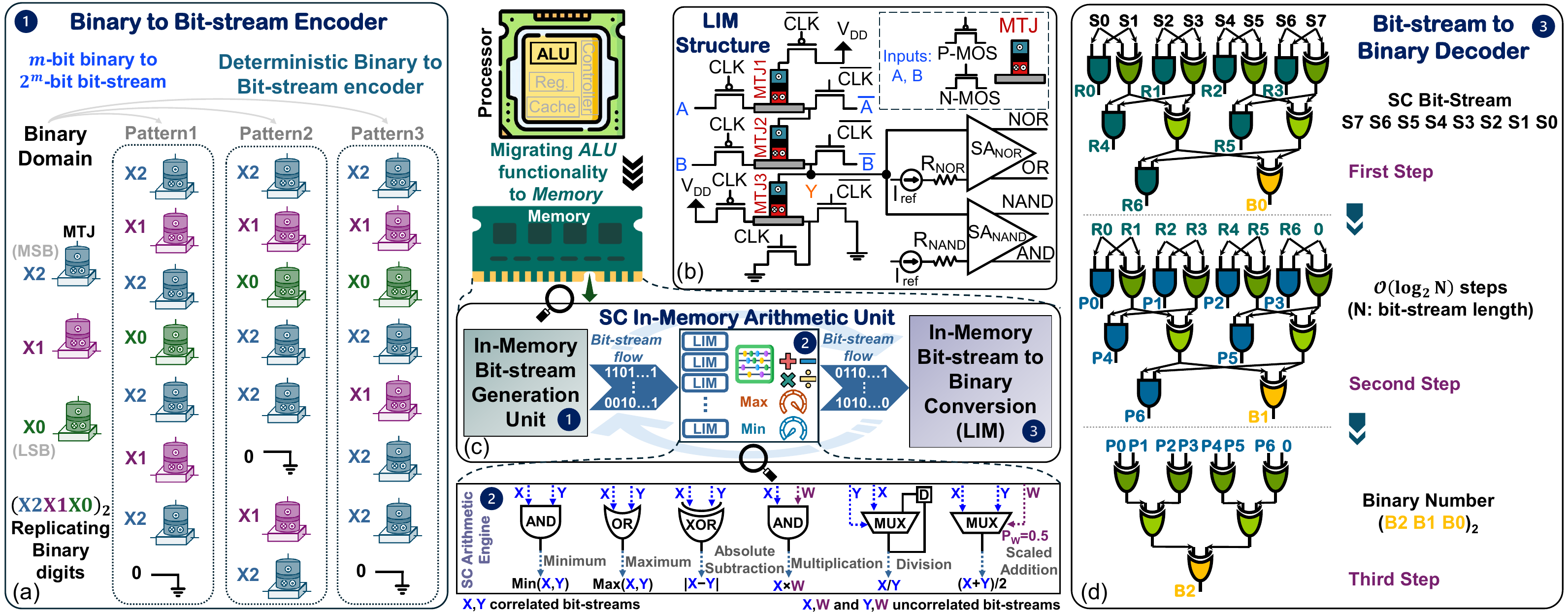}
        \vspace{-1.5em}
        \caption{An overview of \Design, (a) Binary to Bit-stream Encoder, (b) LIM structure, (c) In-Memory ALU engine, and (d) Bit-stream to Binary Decoder.}
        \label{Overall_ALU2}
        \vspace{-0.5em}
\end{figure*}

Motivated by these advantages, 
%Building upon these strengths, 
this work proposes \Design, a robust MTJ-based stochastic IMC %arithmetic 
framework that supports all essential arithmetic operations, mirroring the Arithmetic Logic Unit (ALU) functionality of a general-purpose processor. %addition, subtraction, multiplication, division, minimum, and maximum, supported by the unifies bit-stream generator. 
%The resulting architecture 
\Design~is energy-efficient and lightweight, yet intrinsically resilient to faults and device non-idealities. It %\Design~
supports accurate computations by processing deterministically generated stochastic bit-streams.
Extensive simulations demonstrate %that %the design  
%\Design~maintains 
stable and accurate operation under low supply voltages, significant process variations, and substantial device noise, making \Design~a %. This %These characteristics 
%position the proposed IMC architecture %a 
strong candidate for reliability-critical edge AI applications.
%including edge-AI platforms, ultra-low-power computing nodes, and next-generation IMC-centric accelerators.
Fig.~\ref{SC_binary_comp_white_chart}(b) visualizes the landscape of state-of-the-art IMC designs, illustrating where \Design{} is positioned in terms of \textit{energy consumption} and \textit{robustness}.
The key contributions of this work are as follows:

\noindent\textcolor{Navy}{\ding{172}} An end-to-end MTJ-based stochastic IMC architecture that integrates bit-stream generation, a reconfigurable in-memory arithmetic unit, and bit-stream-to-binary conversion, supporting multiplication, scaled addition, division, absolute subtraction, minimum, and maximum operations entirely within memory.

\noindent\textcolor{Navy}{\ding{173}} A deterministic low-discrepancy (LD) bit mapping mechanism that generates exact stochastic representations with explicit correlation control, eliminating area- and power-intensive random number generation circuitry and switching probability calibration.

\noindent\textcolor{Navy}{\ding{174}} A comprehensive circuit-level reliability evaluation under soft error injection, process variation, and voltage scaling, demonstrating correct operation at noise levels of up to 30\%.

\noindent\textcolor{Navy}{\ding{175}} An application-level validation through morphological image closing, demonstrating robust image processing accuracy under severe device noise and an estimated $10\times$ energy reduction over a von Neumann baseline.

%\Mehcolor{Fig.~\ref{SC_binary_comp_white_chart}(b) visualizes the landscape of state-of-the-art IMC designs, illustrating how \Design{} positions across \textit{energy efficiency} and \textit{robustness}. %occupies the critical white space of \textit{robustness} vs. \textit{energy efficiency}.}

%\vspace{-0.75em}
\section{\Design~Architecture}
\label{prop}
%\vspace{-0.2em}
%proposed method introduces 
\Design~is a reliability-enhanced in-memory %SC 
architecture that embeds the complete arithmetic workflow directly inside MTJ-based memory arrays. As illustrated in Fig.~\ref{Overall_ALU2}, the architecture integrates three primary functional layers: \textcolor{Navy}{\ding{202}} a bit-stream generation unit based on a deterministic bit mapping scheme in MTJ cells, \textcolor{Navy}{\ding{203}} an SC-based IMC arithmetic unit, %reconfigurable 
%IMC computational unit, 
and \textcolor{Navy}{\ding{204}} a bit-stream-to-binary conversion unit. Together, these %components 
form a lightweight and robust %fault-tolerant 
 SC engine that executes a full suite of arithmetic operations in memory without transferring data to external logic units. By confining both data and computation within memory, \Design~achieves high reliability under process variations, voltage scaling, and ambient noise. % conditions.

At the front-end, %front end, 
a deterministic bit mapping scheme %is employed to 
transforms conventional binary inputs into stochastic bit-streams (Fig.~\ref{Overall_ALU2}(a)), without relying on random number generators, which conventionally dominate the area and power cost of SC systems. Our bit mapping utilizes low-discrepancy patterns~\cite{Asadi_FSM_SBS_DATE21} to place each binary digit's contribution at specific positions within the bit-stream, producing exact bit-stream representations and explicit control over inter-stream cross-correlation. 
%distribute the effect of each binary digit across the bit-stream in a controlled and uniform fashion, enabling consistent stochastic representations and controlled correlation behavior in the proposed architecture.
%In particular, 
For an $n$-bit binary input $X = (X_{n-1}, \dots, X_0)$, bit $X_i$ is replicated $2^i$ times and placed at predefined positions specified by one of the LD patterns (e.g., Patterns 1 through 3 in Fig.~\ref{Overall_ALU2}(a) within a bit-stream of length $2^n$). The resulting bit-stream therefore contains a number of logical `1's exactly proportional to the numerical value of the input, providing a deterministic and accurate probability representation free from sampling-induced variation.
Unlike conventional SC generators that rely on pseudo-random sequences or device-level randomness, this scheme removes the need for dedicated random number generator (RNG) circuitry. %hardware and eliminates uncertainty associated with random sampling.
%Furthermore, the correlation between multiple bit-streams can be explicitly controlled through the selection of mapping patterns. 
The choice of LD pattern determines how multiple bit-streams relate to one another: applying the same pattern to two inputs yields correlated %fully cross-correlated 
streams, which are required for operations such as minimum (using bit-wise \texttt{AND}), maximum (using bit-wise \texttt{OR}), and absolute subtraction (using bit-wise \texttt{XOR}), while applying distinct patterns (e.g., Patterns 1 and 2 in Fig.~\ref{Overall_ALU2}(a)) produces effectively independent streams suitable for operations such as multiplication (using bit-wise \texttt{AND}). %and division. 
%This deterministic bit-stream generation approach 
The scheme also integrates naturally with MTJ-based memory arrays:
%enables precise and repeatable stochastic representations while remaining compatible with in-memory implementation. In particular, it integrates naturally with MTJ-based memory arrays, where 
bit replication and placement reduce to %can be realized through 
controlled read and write operations on existing memory cells, eliminating the peripheral logic and area-power overhead that conventional SC bit-stream generators incur.
%thereby avoiding additional peripheral overhead and improving overall energy efficiency.

\begin{comment}
At the front end, a deterministic bit mapping mechanism converts conventional binary inputs to stochastic bit-streams.
%of the system, MTJ-based  random number generators encode numerical values as probabilities and produce stochastic bit-streams that . 
This mapping employs low-discrepancy %, deterministic  
patterns~\cite{Asadi_FSM_SBS_DATE21} that assign individual binary digits to specific positions within the bit-stream. %\red{As illustrated in Fig.~\ref{Overall_ALU2}, a 3-bit binary number stored in memory is converted into an 8-bit SC bit-stream whose density of ones represents the original binary value. According to this mapping scheme, each binary bit $X_i$ is replicated $2^i$ times within the sequence and placed at predefined positions implemented through hardwiring.} 
By selecting same mapping patterns or distinct ones, the system can generate \textit{correlated} or \textit{independent} bit-streams, respectively.
\end{comment}

Following bit-stream generation, the core computation unit performs arithmetic directly within the memory array using a logic-in-memory (LIM) structure~\cite{razi2022toward} (Fig.~\ref{Overall_ALU2}(b)).
%leverages an integrated logic-in-memory (LIM) structure~\cite{razi2022toward} in conjunction with SC logic to execute various arithmetic operations directly within the memory array.
Originally proposed for Boolean operations such as \texttt{AND} and \texttt{OR} inside MTJ arrays, this LIM structure is repurposed here to implement various logic primitives required for SC arithmetic, including \texttt{AND}, \texttt{OR}, \texttt{XOR}, and \texttt{MUX}.
%LIM is an in-memory logic structure originally proposed to realize Boolean operations such as \texttt{AND} and \texttt{OR}. In this work, we utilize this LIM structure to implement the full set of logic functions, including AND, OR, XOR, and MUX, which serve as the building blocks for stochastic arithmetic.
Each LIM cell consists of two MTJs and peripheral FinFET transistors configured to support 
reconfigurable logic functionality. 
As illustrated in Fig.~\ref{Overall_ALU2}(b), inputs $A$ and $B$, represented as stochastic bit-streams generated by the front-end unit (Fig.~\ref{Overall_ALU2}(a)), are applied to the LIM cell. During the \textit{preparation phase}, bidirectional write currents driven by these bit-stream inputs program the magnetic states of the two MTJs %Depending on the input combination, the MTJs are driven 
into parallel (low resistance, $\mathit{L}$) or antiparallel (high resistance, $\mathit{H}$) states, resulting in one of four possible resistance configurations: $\mathit{H}\mathit{H}$, $\mathit{H}\mathit{L}$, $\mathit{L}\mathit{H}$, $\mathit{L}\mathit{L}$.
In the \textit{evaluation phase}, a read current is applied across the MTJ stack. The resulting output voltage is determined through a resistive voltage division at node $Y$, formed between the equivalent resistance of the two programmed MTJs (MTJ1 \& MTJ2) and a fixed reference MTJ (MTJ3). This voltage is then sensed by a dual-output sense amplifier that produces %which resolves the signal into 
complementary logic outputs (e.g., \texttt{AND}/\texttt{NAND} or \texttt{OR}/\texttt{NOR}).
%Because the same cell can be programmed to evaluate various primitives, a single LIM array can be reconfigured %at the bit-stream level 
%to realize the full set of SC arithmetic operations.
By combining this primitive-level reconfigurability with the controlled bit-stream correlation provided by the front-end LD patterns, a single LIM array can be reconfigured to realize the full set of SC arithmetic operations.

\begin{table}[!t]
\centering
\caption{MAE (\%) Comparison of SC Arithmetic Operations%\\ at Various Noise Ratios
}
\vspace{-0.75em}
\resizebox{\columnwidth}{!}{
\begin{tabular}{l c *{7}{c}}
\toprule
% Table Header
\multirow{2}{*}{\begin{tabular}[l]{@{}l@{}}\textbf{SC} \textbf{Arithmetic}\\ \textbf{Operation} \end{tabular}} & \multirow{2}{*}{\begin{tabular}[c]{@{}c@{}}\textbf{Bit-stream}\\ \textbf{Length($N$)} \end{tabular}} & \multicolumn{7}{c}{\textbf{Injected Noise (\%)}} \\
\cmidrule(lr){3-9} % Partial rule under the main header
& & \textbf{0} & \textbf{1} & \textbf{2} & \textbf{5} & \textbf{10} & \textbf{20} & \textbf{30} \\
\midrule

% Data Rows
Multiplication & \multirow{6}{*}{16} & 6.97 & 7.14 & 7.32 & 8.18 & 9.88 & 14.5 & 19.6 \\
Scaled Addition &  & 9.11 & 9.01 & 9.14 & 9.43 & 10.1 & 11.8 & 14.0 \\
Abs. Subtraction &  & 7.88 & 8.03 & 8.05 & 8.79 & 9.94 & 13.4 & 17.5 \\
Division &  & 12.2 & 12.2 & 12.1 & 12.2 & 12.7 & 15.2 & 18.5 \\
Minimum &  & 7.93 & 8.04 & 8.08 & 8.64 & 9.94 & 13.4 & 17.5 \\
Maximum &  & 7.87 & 7.99 & 8.16 & 8.61 & 9.90 & 13.4 & 17.5 \\
\midrule

Multiplication & \multirow{6}{*}{64} & 3.43 & 3.57 & 3.87 & 4.88 & 7.14 & 12.7 & 18.4 \\
Scaled Addition &  & 4.46 & 4.53 & 4.60 & 4.92 & 5.81 & 8.29 & 11.3 \\
Abs. Subtraction &  & 3.92 & 4.03 & 4.16 & 4.87 & 6.56 & 10.9 & 15.7 \\
Division &  & 6.15 & 6.16 & 6.18 & 6.72 & 7.94 & 11.5 & 15.9 \\
Minimum &  & 3.95 & 4.03 & 4.14 & 4.93 & 6.60 & 11.0 & 15.6 \\
Maximum &  & 3.92 & 3.99 & 4.19 & 4.91 & 6.59 & 10.9 & 15.6 \\
\midrule

Multiplication & \multirow{6}{*}{256} & 1.71 & 1.87 & 2.22 & 3.58 & 6.35 & 12.2 & 17.9 \\
Scaled Addition &  & 2.24 & 2.27 & 2.37 & 2.88 & 4.10 & 7.06 & 10.3 \\
Abs. Subtraction &  & 1.96 & 2.07 & 2.27 & 3.30 & 5.49 & 10.2 & 15.3 \\
Division &  & 3.06 & 3.11 & 3.26 & 3.99 & 5.97 & 10.4 & 15.3 \\
Minimum &  & 1.96 & 2.06 & 2.28 & 3.28 & 5.51 & 10.3 & 15.2 \\
Maximum &  & 1.94 & 2.06 & 2.26 & 3.29 & 5.45 & 10.3 & 15.1 \\
\midrule

Multiplication & \multirow{6}{*}{1024} & 0.86 & 1.10 & 1.54 & 3.14 & 6.00 & 11.9 & 17.8 \\
Scaled Addition &  & 1.11 & 1.18 & 1.33 & 2.03 & 3.59 & 6.86 & 10.1 \\
Abs. Subtraction &  & 0.98 & 1.13 & 1.44 & 2.72 & 5.10 & 10.1 & 15.0 \\
Division &  & 1.54 & 1.64 & 1.88 & 3.04 & 5.27 & 10.1 & 15.1 \\
Minimum &  & 0.96 & 1.12 & 1.45 & 2.75 & 5.10 & 10.0 & 14.9 \\
Maximum &  & 0.96 & 1.12 & 1.44 & 2.72 & 5.17 & 10.0 & 15.2 \\
%\midrule
%\begin{tabular}[c]{@{}l@{}} *** \end{tabular} &  &  & 1.1 & 1.6 & 3.0 & 5.3 & 10.2 & 15.1 \\
%\begin{tabular}[c]{@{}l@{}} *** \end{tabular} &  &  & 1.4 & 2.6 & 6.0 & 10.8 & 17.7 & 21.8 \\
\bottomrule
\end{tabular}
}
\label{tab:noise_injection}
\vspace{-0.75em}
\end{table}

%\red{%By %controlling correlation between input bit-streams, %configuring input correlations and control bit-streams, 
%the same LIM structure can be reused to realize a different %higher-level 
%SC operation. %Embedding this reconfigurable logic directly within the memory array enables true in-situ computation, eliminating long data transfers and reducing susceptibility to noise-prone interconnects, while preserving the inherent fault tolerance of stochastic representations.}

Fig.~\ref{Overall_ALU2}(c) demonstrates the in-memory ALU engine of the \Design{} architecture, realized through SC logic primitives. Multiplication is realized %\Design~by %applying 
by bit-wise \texttt{AND}ing %two 
\textit{independent} %uncorrelated 
%stochastic 
bit-streams %to an AND gate
directly inside the memory array. %, producing an output whose probability equals the product of the inputs. 
Minimum and maximum operations are implemented by bit-wise \texttt{AND} and \texttt{OR} logic, respectively, 
%arise naturally by applying 
on \textit{correlated} inputs. %to ensure correct behavior. %to AND or OR logic, respectively, 
%leveraging MTJ-controlled correlation to produce reliable selection behavior. 
Absolute value subtraction is implemented via \texttt{XOR} logic, also on correlated streams. %, enabling robust absolute-difference computation. 
Scaled addition/subtraction and division are carried out using a 2:1 multiplexer (\texttt{MUX}) driven by a control bit-stream. %where the LIM-based \texttt{MUX}  %structure 
%effectively computes weighted averaging or ratios of the input probabilities. 
Because each operation is expressed through simple logic primitives on equally weighted bit-streams, individual bit errors caused by noise or timing irregularities have only a negligible impact on the numerical outcome, which is averaged over the full bit-stream length.

At the back-end of the computation, the resulting bit-stream can either be %directly 
reused as the input to subsequent SC operations or converted back to weighted binary form using an in-memory \textit{bit-stream-to-binary} converter implemented with LIM~\cite{Alam_SC_Memristor_DT_2021} as illustrated in Fig.~\ref{Overall_ALU2}(d). Together, the MTJ-driven bit-stream generator, the reconfigurable LIM computational fabric, and the inherently resilient SC operation set form an end-to-end, highly robust in-memory arithmetic architecture. %By capitalizing on the statistical nature of SC and the physical robustness of MTJs, the system maintains correct functionality even under significant process variations and aggressively scaled voltages. This reliability-centric design positions the architecture as a suitable candidate for future edge AI, low-power sensing, and IMC-centric accelerator platforms where robust operation is mandatory despite limited hardware overhead and unpredictable operating conditions.
%, tightly integrated,

\color{black}
\section{Experimental Results}
\label{exp}

\subsection{Performance Evaluations}

\begin{table}[!t]
\centering
%\setlength{\tabcolsep}{4pt}
%\renewcommand{\arraystretch}{0.6}
% \relscale{0.88}
\caption{Performance Evaluation of SC Arithmetic Operations}
\vspace{-0.75em} 
 \resizebox{1.0\columnwidth}{!}{
\begin{tabular}{lcccc} 
\toprule
\begin{tabular}[l]{@{}l@{}}\textbf{SC Arithmetic}\\\textbf{Operations}\end{tabular} & \begin{tabular}[c]{@{}c@{}} \textbf{Preparation} \\\textbf{Delay}($ns$) \end{tabular} & \begin{tabular}[c]{@{}c@{}} \textbf{Evaluation}\\ \textbf{Delay}($ps$) \end{tabular} & \begin{tabular}[c]{@{}c@{}} \textbf{Power}\\ ($\mu W$) \end{tabular}& \begin{tabular}[c]{@{}c@{}} \textbf{Cycles}\\ \textbf{Count} \end{tabular} \\ 
\midrule
Scaled Addition & 1.79 & 3.41 & 144.5 & N$+0.5$ \\
%\hline
Abs. Subtraction & 1.72 & 3.49 & 144.8 & N$+0.5$  \\
%\hline
Multiplication \& Minimum & 1.17 & 5.34 & 48.4 & N  \\
%\hline
Division & 1.72 & 3.53 & 145.5 & N$+0.5$  \\
%\hline
Maximum & 1.17 & 3.37 & 48.4 & N  \\
\bottomrule
\end{tabular}
}
\label{tab:SimulationResults}
%\vspace{-0.5em}
\end{table}

To evaluate the robustness of the underlying SC-based arithmetic units in \Design{}, we injected controlled ratios of soft errors into the primitive logic blocks implementing different arithmetic operations. %As can be seen in 
Table~\ref{tab:noise_injection} reveals two trends: \ding{172} increasing the stochastic bit-stream length
improves both computation accuracy and noise tolerance, 
%consistently improves the computation accuracy, % overall performance
and \ding{173} \Design~maintains high tolerance to substantial soft error rates %extensive soft errors 
while preserving acceptable accuracy. %levels.
%\ding{174} 
%Although a bit-stream length of 1024 yields the highest accuracy among the reported results, a length of 256 achieves sufficiently low error for practical applications. Because SC processing latency scales linearly with bit-stream length, this shorter length offers a favorable trade-off between accuracy and latency.
Table~\ref{tab:noise_injection} confirms that increasing bit-stream length monotonically improves both accuracy and noise tolerance across all operations. The appropriate length is therefore application-dependent; latency-sensitive tasks may favor shorter streams, while applications demanding higher noise margins benefit from longer streams.

%These observations reinforce that the proposed architecture delivers dependable, energy-efficient operation, well-suited for variation-sensitive environments.
\begin{figure}[!t]
%\color{teal}
    \centering
        \includegraphics[width=\linewidth]{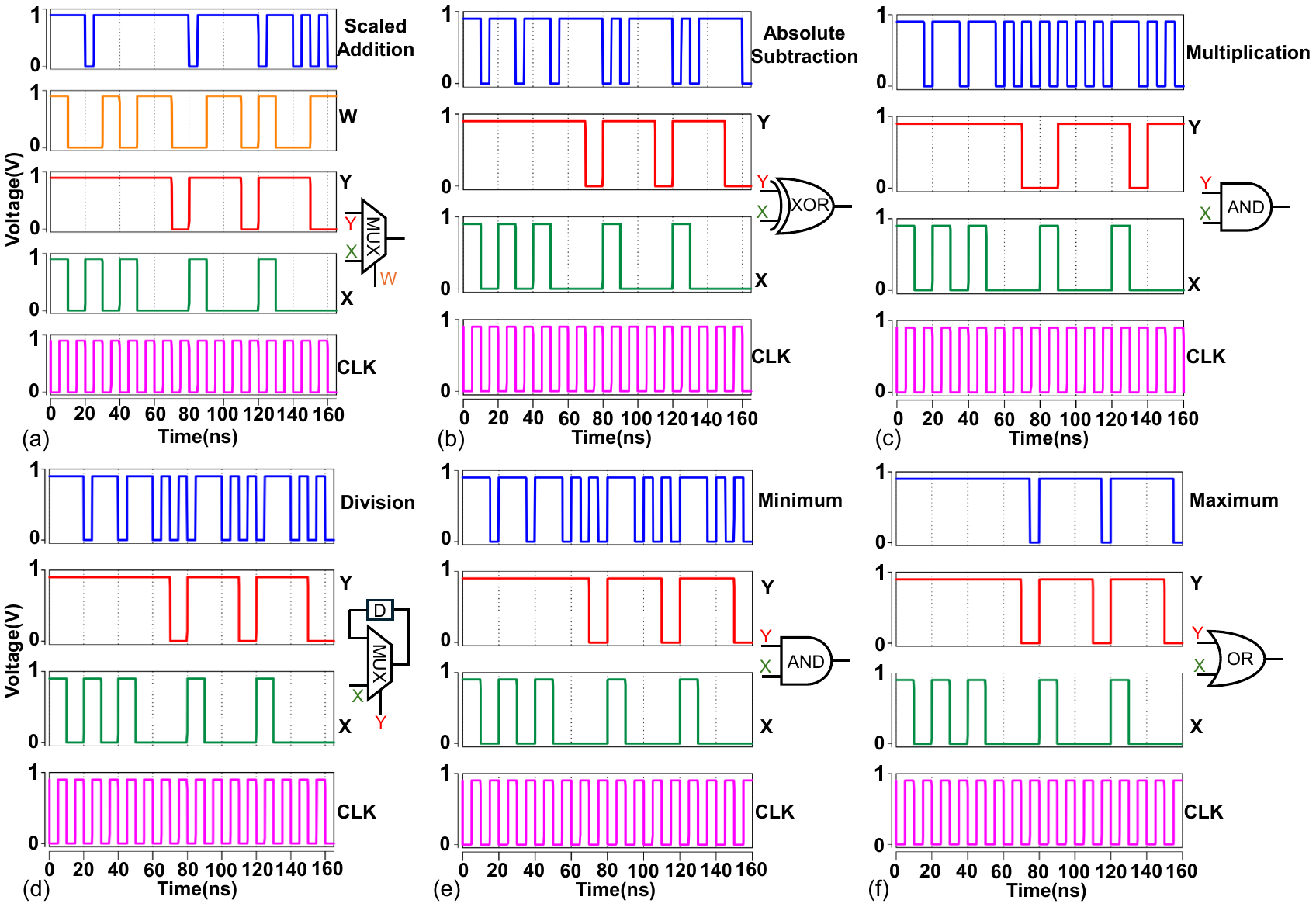}
        \vspace{-1.75em}
        \caption{Timing diagram of SC arithmetics, (a)~Scaled Addition, (b)~Absolute Subtraction, (c)~Multiplication, (d)~Division, (e)~Minimum, and (f)~Maximum.}
        \label{fig:plots_operations}
        \vspace{-1.0em}
\end{figure}

%---------------------------------------------
%           Plots
%---------------------------------------------
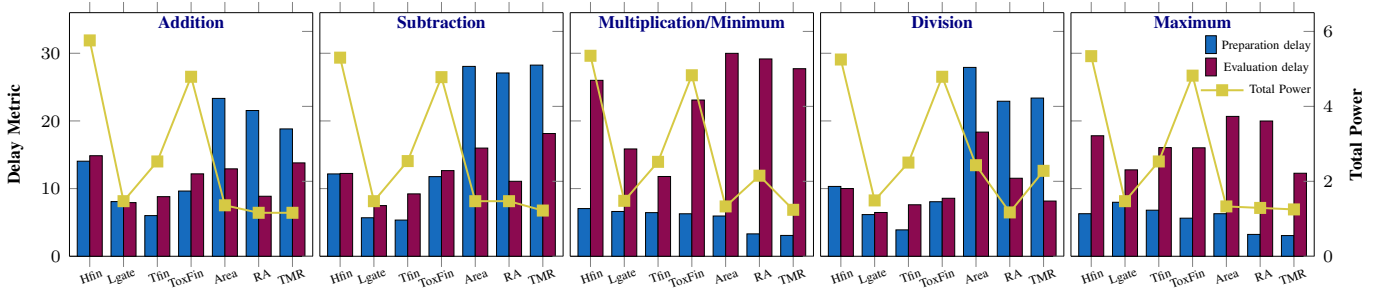
\begin{figure*}[!t]
\centering
% This explicitly scales the entire plot to fit exactly within the page margins
\resizebox{\textwidth}{!}{%
\begin{tikzpicture}

% =========================================================
% GROUP 1: Left Y-Axis for Delay Metrics (Bar Charts)
% =========================================================
\begin{groupplot}[
    group style={
        group size=5 by 1,
        horizontal sep=0.1cm, % space between plots
        y descriptions at=edge left,
    },
    width=5cm, % Absolute width won't break margins now due to resizebox
    height=5cm,
    ymin=0, ymax=36, % Keeps title and legend clear of the bars
    enlarge x limits=0.1,
    symbolic x coords={Hfin, Lgate, Tfin, ToxFin, Area, RA, TMR},
    xtick=data,
    x tick label style={align=center, xshift=8pt, yshift=-0.8pt, rotate=20, anchor=east, font=\tiny},
    ylabel={Delay Metric},
    y label style={font=\scriptsize\bfseries},
    tick label style={font=\scriptsize},
    axis y line*=left, % Bind to left y-axis
    title style={at={(0.5, 0.96)}, anchor=north}, 
]

% Plot 1: ADD
\nextgroupplot[ybar=0pt, bar width=5pt, title={\textcolor{Navy}{\scriptsize\textbf{Addition}}}]
\addplot[fill=DodgerBlue2!80!black, draw=black] coordinates {(Hfin,14.05) (Lgate,8.08) (Tfin,6.01) (ToxFin,9.64) (Area,23.33) (RA,21.54) (TMR,18.82)};
\addplot[fill=DeepPink4!100!black, draw=black] coordinates {(Hfin,14.86) (Lgate,7.93) (Tfin,8.81) (ToxFin,12.18) (Area,12.92) (RA,8.87) (TMR,13.80)};

% Plot 2: SUB
\nextgroupplot[ybar=0pt, bar width=5pt, title={\textcolor{Navy}{\scriptsize\textbf{Subtraction}}}]
\addplot[fill=DodgerBlue2!80!black, draw=black] coordinates {(Hfin,12.17) (Lgate,5.69) (Tfin,5.35) (ToxFin,11.77) (Area,28.06) (RA,27.09) (TMR,28.25)};
\addplot[fill=DeepPink4!100!black, draw=black] coordinates {(Hfin,12.24) (Lgate,7.49) (Tfin,9.22) (ToxFin,12.66) (Area,15.99) (RA,11.08) (TMR,18.14)};

% Plot 3: MUL/MIN
\nextgroupplot[ybar=0pt, bar width=5pt, title={\textcolor{Navy}{\scriptsize\textbf{Multiplication/Minimum}}}]
\addplot[fill=DodgerBlue2!80!black, draw=black] coordinates {(Hfin,7.05) (Lgate,6.62) (Tfin,6.45) (ToxFin,6.28) (Area,5.95) (RA,3.31) (TMR,3.08)};
\addplot[fill=DeepPink4!100!black, draw=black] coordinates {(Hfin,26.00) (Lgate,15.86) (Tfin,11.79) (ToxFin,23.09) (Area,29.99) (RA,29.16) (TMR,27.72)};

% Plot 4: DIV
\nextgroupplot[ybar=0pt, bar width=5pt, title={\textcolor{Navy}{\scriptsize\textbf{Division}}}]
\addplot[fill=DodgerBlue2!80!black, draw=black] coordinates {(Hfin,10.32) (Lgate,6.16) (Tfin,3.90) (ToxFin,8.06) (Area,27.91) (RA,22.91) (TMR,23.38)};
\addplot[fill=DeepPink4!100!black, draw=black] coordinates {(Hfin,10.01) (Lgate,6.47) (Tfin,7.61) (ToxFin,8.57) (Area,18.35) (RA,11.53) (TMR,8.16)};

% Plot 5: MAX
\nextgroupplot[
    ybar=0pt, 
    bar width=5pt, 
    title={\textcolor{Navy}{\scriptsize\textbf{Maximum}}},
    legend style={at={(1.04, 0.93)}, anchor=north east, legend columns=1, draw=none, fill=none, font=\tiny},
    legend image code/.code={ % legend inside the plot
        \draw[#1] (0cm,-0.1cm) rectangle (0.15cm,0.15cm);
    }
]
\addplot[fill=DodgerBlue2!80!black, draw=black] coordinates {(Hfin,6.29) (Lgate,7.98) (Tfin,6.81) (ToxFin,5.63) (Area,6.29) (RA,3.23) (TMR,3.06)};
\addplot[fill=DeepPink4!100!black, draw=black] coordinates {(Hfin,17.81) (Lgate,12.77) (Tfin,16.06) (ToxFin,16.03) (Area,20.67) (RA,19.99) (TMR,12.27)};
\legend{Preparation delay, Evaluation delay}

\end{groupplot}

% =========================================================
% GROUP 2: Right Y-Axis for Power Metric (Line Charts)
% =========================================================
\begin{groupplot}[
    group style={
        group size=5 by 1,
        horizontal sep=0.1cm, % space between plots
        y descriptions at=edge right,
    },
    width=5cm,
    height=5cm,
    ymin=0, ymax=6.5, 
    enlarge x limits=0.1,
    symbolic x coords={Hfin, Lgate, Tfin, ToxFin, Area, RA, TMR},
    axis x line=none, % Hide x-axis to prevent double drawing
    axis y line*=right, % Bind to right y-axis
    ylabel={Total Power},
    y label style={font=\scriptsize\bfseries},
    tick label style={font=\scriptsize},
    % GLOBAL LEGEND REMOVED FROM HERE
]

% Plot 1: ADD Power
\nextgroupplot
\addplot[color=yellow!80!black, mark=square*, thick] coordinates {(Hfin,5.76) (Lgate,1.47) (Tfin,2.53) (ToxFin,4.79) (Area,1.36) (RA,1.16) (TMR,1.16)};

% Plot 2: SUB Power
\nextgroupplot
\addplot[color=yellow!80!black, mark=square*, thick] coordinates {(Hfin,5.30) (Lgate,1.47) (Tfin,2.54) (ToxFin,4.78) (Area,1.47) (RA,1.47) (TMR,1.22)};

% Plot 3: MUL/MIN Power
\nextgroupplot
\addplot[color=yellow!80!black, mark=square*, thick] coordinates {(Hfin,5.35) (Lgate,1.48) (Tfin,2.52) (ToxFin,4.83) (Area,1.33) (RA,2.15) (TMR,1.24)};

% Plot 4: DIV Power
\nextgroupplot
\addplot[color=yellow!80!black, mark=square*, thick] coordinates {(Hfin,5.25) (Lgate,1.49) (Tfin,2.50) (ToxFin,4.79) (Area,2.43) (RA,1.17) (TMR,2.28)};

% Plot 5: MAX Power
% LEGEND MOVED HERE: Placed right below the delay legend
\nextgroupplot[
    legend style={at={(1.04, 0.75)}, anchor=north east, legend columns=1, draw=none, fill=none, font=\tiny}
]
\addplot[color=yellow!80!black, mark=square*, thick] coordinates {(Hfin,5.34) (Lgate,1.47) (Tfin,2.53) (ToxFin,4.82) (Area,1.33) (RA,1.29) (TMR,1.25)};
\legend{Total Power}

\end{groupplot}
\end{tikzpicture}%
} % End of resizebox
\vspace{-2.5em}
\caption{
Impact of process variation in preparation delay, evaluation delay, and power consumption for various arithmetic operations in \Design{}.
%Maximum deviation of \Design{} in preparation delay, evaluation delay, and total power consumption under process variation. %Delay components are plotted as bars on the primary (left) y-axis, while total power is represented as a line on the secondary (right) y-axis.
}
\label{fig:pv_plots}
%\vspace{-1.0em}
\end{figure*}

%To further validate the resilience and functional correctness, %of the proposed IMC framework, %in-memory stochastic arithmetic architecture,
To complement the behavioral robustness analysis with circuit-level validation,
we conducted comprehensive HSPICE simulations using a 14\,$nm$ FinFET technology model for the peripheral circuitry. %along with a compact device model for in-plane anisotropy SHE-MTJs~\cite{kang2016spintronic}, following the simulation setup described in~\cite{razi2022toward}. 
Fig.~\ref{fig:plots_operations}(a)--(f) show the transient responses of representative arithmetic operations and validate the correct temporal behavior of the two-phase LIM-based computation: input programming in the preparation phase, followed by resistance-based sensing during the evaluation, produces the expected logical output for every operation.
Table~\ref{tab:SimulationResults} reports the corresponding  preparation delay, evaluation delay, and power consumption, which vary with the underlying logic primitive. %The main performance metrics extracted from these simulations are summarized in Table~\ref{tab:SimulationResults}. 
%The preparation and evaluation delays, as well as total power consumption, vary across operations depending on the underlying logic primitive. 
\texttt{AND}-based operations (multiplication and minimum), exhibit the shortest delays and lowest power owing to their simpler structure, while \texttt{MUX}-based operations (scaled addition and division) incur higher delays %preparation and evaluation times 
but remain well within practical limits for stochastic arithmetic. 
%As expected of a serial SC design, 
Since \Design~processes bit-streams serially (one bit per cycle), the cycle count grows linearly with the bit-stream length $N$ for all operations.
%As expected, the number of processing cycles scales linearly with the bit-stream length for all operations. The proposed design follows a serial SC design, where latency scales linearly with the bit-stream length $N$.
However, unlike conventional IMC approaches that require a long chain of operations %and a large number of sequential operations 
to realize arithmetic functions such as multiplication and division, 
SC realizes these operations using simple logic primitives, resulting in competitive or reduced effective latency for complex computations. 
Collectively, these results verify that \Design~performs arithmetic operations reliably within the memory array while maintaining a low hardware footprint.

To assess robustness against fabrication and device-level uncertainties, we performed a process variation analysis %was carried out 
using Monte Carlo simulation. A $\pm 3\sigma$ Gaussian distribution with 10\% variation was applied to critical FinFET parameters, namely gate length (Lgate), fin height (Hfin), fin thickness (Tfin), and oxide thickness (ToxFin), as well as to the key MTJ characteristics such as free-layer area (Area), tunnel magnetoresistance (TMR), and resistance--area product (RA). Fig.~\ref{fig:pv_plots} reports the maximum deviation in preparation delay, evaluation delay, and total power across all operations under this variation. %As it can be seen, process variations increase total power consumption, preparation and evaluation delays across operations.
While each metric shows measurable spread, 
%Nevertheless, Monte Carlo simulations confirm that 
the functional outputs of every arithmetic operation remain correct,
confirming that \Design{} tolerates realistic device-level variations without loss of computational integrity.
%demonstrating strong tolerance to device-level variations. 

We further examined operational stability %conducted a complementary evaluation to examine
%A complementary evaluation examined 
%system behavior and its operational stability 
under supply voltage scaling. %to determine %both 
%its %energy efficiency and 
%operational stability. 
Simulations were performed at the nominal 0.9\,V, as well as at 0.8\,V, and 0.7\,V. %and then at 0.8\,V and 0.7\,V. 
All operations maintained correct functionality across the three voltage levels, 
%Across all three voltage levels, the implemented %stochastic arithmetic 
%operations maintained correct functionality without %any 
with no measurable degradation in accuracy. Lowering the supply voltage yielded substantial power consumption savings relative to the nominal point: %with savings of 
approximately 27\% at 0.8\,V and 51\% at 0.7\,V. %, while preserving the expected outputs. 
This robustness %consistent behavior 
to %aggressive 
voltage scaling stems from the stochastic representation itself: %underscores the inherent reliability of the implemented architecture: %the in-memory computing structure, 
voltage-induced variations in MTJ write and read currents perturb individual bit values but leave the statistical distribution of `1's in each bit-stream essentially unchanged, preserving the encoded numerical value.
%do not significantly impact the statistical behavior of the resulting bit-streams. 

\begin{comment}
\begin{figure}[!t]
    \centering
        \includegraphics[width=\linewidth]{Figures/PV_plots2.pdf}
        \vspace{-1em}
        \caption{Maximum deviation in preparation delay, evaluation delay, and total power consumption under process variation.}
        \label{fig:pv_plots}
\end{figure}
\end{comment}

\begin{figure}[!b]
    \centering
        %\vspace{-1.0em}
        \includegraphics[width=\linewidth]{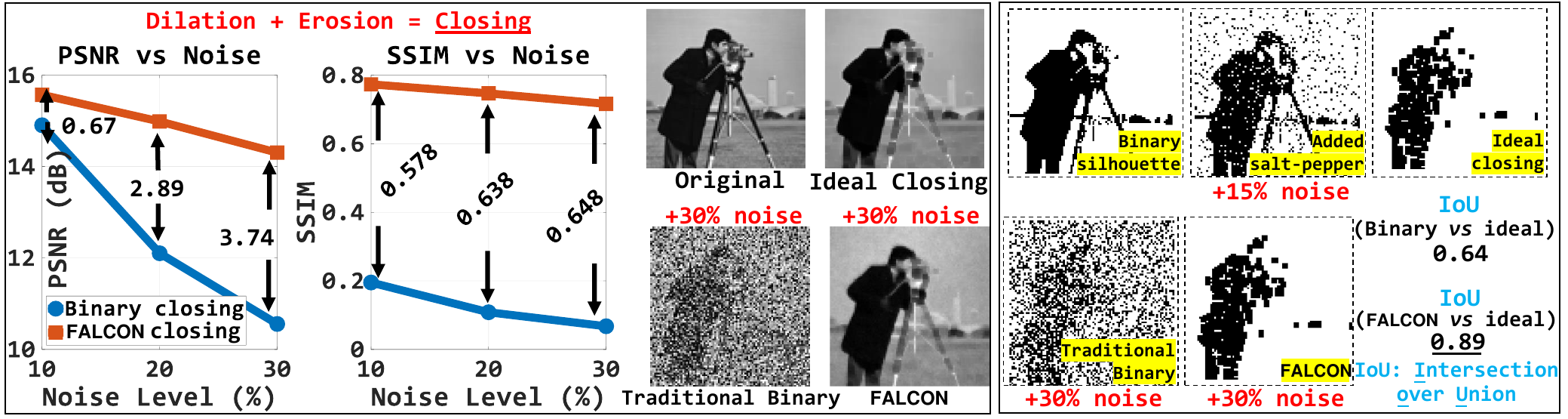}
        \vspace{-2em}
        \caption{Noise robustness of \textit{morphological closing} with \Design~versus conventional binary IMC.
        %Noise-robust \textit{morphological closing} on images via SC.
        %Image \textit{closing} case studey; \Design~ vs. Binary computing.
        }
        \label{app1_label}
        %\vspace{-1.5em}
\end{figure}

\begin{table*}[!t]
%\vspace{-1em}
\centering
%\normalsize
%\renewcommand{\arraystretch}{0.6}
\caption{Comparison Summary of the proposed \Design{} and State-of-The-Art IMC Architectures}
\vspace{-0.75em}
\label{tab:comprehensive_sota_comp}
\setlength{\tabcolsep}{2pt}
\resizebox{\textwidth}{!}{
% The last column is injected with a light blue background for emphasis
\begin{tabular}{@{}lcccccccccc>{\columncolor{Lavender!40}}c@{}} 
\toprule
\textbf{Features} & 
\textbf{\begin{tabular}[c]{@{}c@{}} \textcolor{Navy}{MICRO'17~\cite{Ambit_MICRO2017}} \end{tabular}} & 
\textbf{\begin{tabular}[c]{@{}c@{}} \textcolor{Navy}{ICCAD'19~\cite{ReDRAM_Angizi_ICCAD2019}} \end{tabular}} & 
\textcolor{Navy}{\textbf{DATE'20}~\cite{SCRIMP_Imani_DATE2020}} & 
\textbf{\begin{tabular}[c]{@{}c@{}} \textcolor{Navy}{D\&T'21~\cite{Alam_SC_Memristor_DT_2021}} \end{tabular}} & 
\textbf{\begin{tabular}[c]{@{}c@{}} \textcolor{Navy}{ASPLOS'21~\cite{SIMDRAM_ASPLOS2021}} \end{tabular}} & 
\textbf{\begin{tabular}[c]{@{}c@{}} \textcolor{Navy}{APL'21~\cite{DW_MTJ_APL2021}} \end{tabular}} & 
\textbf{\begin{tabular}[c]{@{}c@{}} \textcolor{Navy}{Access'22~\cite{MagCiM_Access2022}} \end{tabular}} & 
\textbf{\begin{tabular}[c]{@{}c@{}}\textcolor{Navy}{\textbf{JXCDC'23}~\cite{SC_CRAM_2023}}\end{tabular}} & 
\textcolor{Navy}{\textbf{DAC'25~\cite{All_in_Mem_SC_DAC2025}}} & 
\textcolor{Navy}{\textbf{IJEC'25~\cite{hajisadeghi2025stoch}}} & 
\textbf{\begin{tabular}[c]{@{}c@{}}\textcolor{Navy}{\Design}\end{tabular}} \\ \midrule

\textbf{Computation} & 
Binary+IMC & 
Binary+IMC & 
SC+IMC & 
SC+IMC & 
Binary+IMC & 
Binary+IMC & 
Binary+IMC & 
SC+IMC & 
SC+IMC & 
SC+IMC & 
SC+IMC \\ \midrule

\textbf{Technology} & 
Com. DRAM & 
Com. DRAM & 
ReRAM & 
Memristive & 
Com. DRAM & 
SOT-MTJ & 
MTJ (mCell). & 
MTJ (2T1M) & 
ReRAM & 
STT-MRAM & 
MTJ+FinFET \\ 

\midrule

\textbf{Operations} & 
\begin{tabular}[c]{@{}c@{}}Bulk\\ Bitwise\end{tabular} & 
\begin{tabular}[c]{@{}c@{}}Reconfigurable\\ Bitwise\end{tabular} & 
\begin{tabular}[c]{@{}c@{}} General\\ SC Arith. \end{tabular} & 
\begin{tabular}[c]{@{}c@{}}Exact\\ Multiplication\end{tabular} & 
\begin{tabular}[c]{@{}c@{}}Flexible\\ SIMD Arith.\end{tabular} & 
\begin{tabular}[c]{@{}c@{}}Basic Logic\\ Gates\end{tabular} & 
\begin{tabular}[c]{@{}c@{}} Full Arith.\\ Suite (ALU)\end{tabular} & 
\begin{tabular}[c]{@{}c@{}}Neuromorphic\\ Arith.\end{tabular} & 
\begin{tabular}[c]{@{}c@{}} Full Arith.\\ Suite  \end{tabular} & 
\begin{tabular}[c]{@{}c@{}} Bit-Parallel\\ Arith. \end{tabular} & 
\begin{tabular}[c]{@{}c@{}}Full Arith.\\ Suite (ALU)\end{tabular} \\ 

\midrule

{\textbf{Latency}} & 
\begin{tabular}[c]{@{}c@{}}Macro-level row\\ activations\end{tabular} & 
\begin{tabular}[c]{@{}c@{}}DRAM timing\\ constraints ($\sim$100ns)\end{tabular} & 
\begin{tabular}[c]{@{}c@{}} 2 cycles\\ (multiplication) \end{tabular} & 
\begin{tabular}[c]{@{}c@{}} 6 cycles\\(full-precision) \end{tabular} & 
\begin{tabular}[c]{@{}c@{}}Multi-cycle\\ bit-serial ops\end{tabular} & 
\begin{tabular}[c]{@{}c@{}}Sub-nanosecond\\ switching\end{tabular} & 
\begin{tabular}[c]{@{}c@{}} $\sim$1$\mu$s \end{tabular} & 
\begin{tabular}[c]{@{}c@{}} $\mathcal{O}(N)$ cycles \end{tabular} & 
\begin{tabular}[c]{@{}c@{}} 0.08--12.5$\mu$s \end{tabular} & 
\begin{tabular}[c]{@{}c@{}} 0.01--22.5$\times$\\ (Normalized) \end{tabular} & 
\begin{tabular}[c]{@{}c@{}} $\mathcal{O}(N)$ cycles \end{tabular} \\ 

\midrule

\multirow{3}{*}{\begin{tabular}[c]{@{}l@{}} \textbf{Power/Energy}\\ \textbf{Consumption} \end{tabular}} & 
\textbf{High} & 
\textbf{High} & 
\textbf{Low} & 
\textbf{Low} & 
\textbf{High} & 
\textbf{Low} & 
\textbf{Low} & 
\textbf{Low--Medium} & 
\textbf{Low} & 
\textbf{Medium} & 
\textbf{Low} \\

 & 
\begin{tabular}[c]{@{}c@{}}Array act.\\ power\end{tabular} & 
\begin{tabular}[c]{@{}c@{}}Sensing \&\\ row-buffer\end{tabular} & 
\begin{tabular}[c]{@{}c@{}} 45.2fJ/op\\ (multiplication) \end{tabular} & 
\begin{tabular}[c]{@{}c@{}} 90pJ\\(6-bit precision) \end{tabular} & 
\begin{tabular}[c]{@{}c@{}}Macro-level\\ row activations\end{tabular} & 
\begin{tabular}[c]{@{}c@{}}SOT\\ efficiency\end{tabular} & 
\begin{tabular}[c]{@{}c@{}} $\sim$0.08--0.3nJ \end{tabular} & 
\begin{tabular}[c]{@{}c@{}} High\\ efficiency \end{tabular} & 
\begin{tabular}[c]{@{}c@{}} 3.50--4.48nJ \end{tabular} & 
\begin{tabular}[c]{@{}c@{}} 0.98--15.4$\times$\\ (Normalized) \end{tabular} & 
\begin{tabular}[c]{@{}c@{}}48.4--145.5$\mu$W\\ (per op)\end{tabular} \\ 

\midrule

\textbf{Fault Tolerance} & 
\textbf{Low}$\Downarrow$ & 
\textbf{Low}$\Downarrow$ & 
\textbf{High}$\Uparrow$ & 
\textbf{High}$\Uparrow$ & 
\textbf{Low}$\Downarrow$ & 
\textbf{Low}$\Downarrow$ & 
\textbf{Low}$\Downarrow$ & 
\textbf{High}$\Uparrow$ & 
\textbf{High}$\Uparrow$ & 
\textbf{High}$\Uparrow$ & 
\textbf{High}$\Uparrow$ \\ 

\bottomrule

\end{tabular}
}
%\vspace{-1.5em}
\end{table*}

\subsection{Case Study: Morphological Image Closing}
%\vspace{-0.5em}
We further assess our proposed \Design{} architecture on a realistic %noise-tolerant 
edge AI image processing task: \textit{morphological closing} on grayscale and binary data.
Morphological closing %requires sequential 
consists of dilation (local maximum) followed by erosion (local minimum). Rather than transferring pixel arrays to an external processor, \Design{} %natively 
executes these operations entirely in-situ.

First, the image pixel intensities are encoded into bit-streams of length $N$=1024 using \Design’s MTJ-driven deterministic bit-mapping. Because minimum and maximum operations inherently require correlated inputs, \Design{} applies the same LD pattern (see Fig.~\ref{Overall_ALU2}(a)) to  adjacent pixel operands. The computation is then evaluated directly within the memory array using \Design's reconfigurable LIM structures configured for bit-wise \texttt{OR} (Max) and \texttt{AND} (Min) logic.
Fig.~\ref{app1_label} illustrates the resulting robustness under severe device noise. 
%demonstrates how \Design{} enables robust closing under severe device noise. 
We applied a $3\times3$ grayscale closing to a sample image and compared a conventional binary IMC approach with \Design’s stochastic execution as noise injection increases from 10\% to 30\%. 
While the binary closing collapses rapidly, \Design{}'s deterministic SC processing remains largely stable.
We further validated this behavior on a binary silhouette: starting from a clean mask, we added 15\% salt-and-pepper noise and performed a $3\times3$ closing under 30\% device noise. 
The conventional binary execution breaks the silhouette and leaves spurious pixels, yielding an Intersection over Union (IoU) of only 0.64 with the ideal mask. 
By contrast, \Design{} %successfully
recovers a clean, connected shape with an IoU of 0.89. 
These results confirm that \Design{}'s %proposed 
bit mapping and LIM execution maintain robust, accurate system-level behavior even under extreme device-level variations.

Based on circuit-level extractions, executing a $256 \times 256$ morphological closing in-situ costs only 7.63 $\mu$J. Eliminating off-chip data movement entirely, this yields an estimated energy saving of roughly 10$\times$ compared to standard sliding-window executions on a von Neumann baseline, while maintaining robust accuracy under device noise.

%Fig.~\ref{app1_label} shows how our \Design{} Min/Max cells (\texttt{AND}/\texttt{OR}) enable robust closing, i.e., dilation followed by erosion, under severe device noise. We applied 3$\times$3 grayscale closing to the Cameraman image and compared conventional \textit{binary closing} with \textit{SC closing} ($N$=1024) as the noise injection increases from 10\% to 30\%. While \textit{binary closing} collapses rapidly, \textit{SC closing} remains largely stable. The example images at 30\% noise highlight this effect. We also repeat the experiment on a binary silhouette; starting from a clean mask, we add 15\% salt-and-pepper noise and then perform 3$\times$3 closing under 30\% noise. \textit{Binary closing} breaks the silhouette and leaves spurious pixels (IoU with the ideal mask = 0.64). By contrast, \textit{SC closing} recovers a clean, connected shape (IoU = 0.89). \red{This demonstrates that the proposed architecture maintains robust system-level behavior under significant device noise, as reflected by stable application-level metrics (IoU).}
%\red{\ul{[Hassan: A problem with the case study is that it's disconnected from the proposed FALCON. They expect that the case study be used to validate the proposed architecutre.}}

%------------------------------------------------------------

\section{Comparison with State of the Art}

Recent advancements in IMC and emerging non-volatile technologies have aggressively targeted the von Neumann memory wall, though they frequently trade off reliability, arithmetic flexibility, and power consumption againt one another. Binary IMC architectures built on commodity DRAM, such as Ambit~\cite{Ambit_MICRO2017}, ReDRAM~\cite{ReDRAM_Angizi_ICCAD2019}, and %advanced frameworks like 
SIMDRAM~\cite{SIMDRAM_ASPLOS2021}, elevate physical memory primitives into flexible, bit-serial single-instruction multiple-data (SIMD) arithmetic. 
However, %despite these throughput optimizations, 
these DRAM-based computations inevitably incur significant macro-level power overheads due to multi-cycle row activations and remain inherently vulnerable to soft errors. In contrast, localized in-memory logic based on emerging non-volatile devices offers substantial energy efficiency. 
%For example, 
Architectures such as MagCiM~\cite{MagCiM_Access2022} utilize MTJs to execute energy-efficient Boolean logic directly within the memory array, %effectively addressing leakage power and bandwidth constraints. Similarly, 
while Domain Wall MTJ (DW-MTJ) devices~\cite{DW_MTJ_APL2021} leverage spin-orbit torque (SOT) for sub-nanosecond logic switching. 
Yet, because these %flexible 
in-memory processors rely on strict, bit-significant binary representations, they retain the traditional susceptibility to device-level variations and single-event upsets.

Integrating SC into IMC frameworks brings robust fault tolerance, mitigating the need for dedicated error correction mechanisms within memory. 
Frameworks such as SCRIMP~\cite{SCRIMP_Imani_DATE2020}, SC-CRAM~\cite{SC_CRAM_2023}, and Stoch-IMC~\cite{hajisadeghi2025stoch} share the goal of embedding both bit-stream generation and arithmetic within the memory array.
However, they differ fundamentally in execution: they rely on the intrinsic switching stochasticity of the memory device itself (e.g., ReRAM sub-threshold switching or thermal noise-driven MTJ switching) to generate random bit-streams. This probabilistic generation makes cross-correlation management between streams %quite 
challenging.
Other architectures, such as All-in-Memory SC on ReRAM~\cite{All_in_Mem_SC_DAC2025}, take a hybrid approach by decoupling a true random number generator from bit-stream generation, but they still depend on a separately generated random operand and must carefully manage the high energy cost and limited endurance inherent in ReRAM write operations.

\Design{} overcomes these limitations through a structurally different %technological 
approach. By employing a deterministic bit mapping mechanism via standard MTJ read/write operations, \Design{} natively and tightly controls inter-stream cross-correlation. This %vital 
distinction allows \Design{} to sidestep ReRAM endurance limitations and complex global accumulation networks, executing robust operations (including division, absolute subtraction, and min/max) entirely in-situ at the microwatt scale. 
Because \Design{} exploits only deterministic MTJ read/write behavior, it %completely 
eliminates the need for area-intensive RNG circuitry and for the %dependence on 
switching probability calibration required by noise-driven designs. 
Table~\ref{tab:comprehensive_sota_comp} provides a comparison %overview 
of these architectures, highlighting the categorical divide between binary-IMC designs, which achieve high throughput at the cost of vulnerability, and SC-IMC designs, which trade some throughput for intrinsic robustness.

\color{black}

\section{Conclusion}
\label{conc}

This work presented \Design, a fault-tolerant MTJ-based stochastic in-memory computing architecture that integrates deterministic stochastic bit-stream generation with reconfigurable LIM computation. By combining the inherent robustness of stochastic representations with compact MTJ-based in-memory logic, \Design{} executes essential arithmetic operations directly inside memory arrays while minimizing data movement and peripheral overhead. Simulation results demonstrate reliable operation under substantial soft error injection, process variation, and aggressive voltage scaling. 
Beyond circuit-level validation, a morphological image closing case study confirms robust system-level behavior and significant energy savings, underscoring the framework's suitability for energy-constrained, reliability-critical edge-AI applications
%\color{teal}In addition, the proposed framework achieves robust system-level behavior and significant energy savings in a morphological image closing case study, underscoring its suitability for energy-constrained, reliability-critical edge AI applications.
\color{black}
\vspace{-0.5em}

\vspace{-0.5em}
\section*{Acknowledgment}
\vspace{-0.2em}
This work is supported in part by National Science Foundation (NSF) under grants \#2227578, \#2415836, \#2019511, \#2339701, \#2609436, \#2616837 National Aeronautics and Space Administration (NASA) grant 80NSSC25C0335, in part by Vernon and Ruby Langlinais Non-Endowed Research Fund, in part by the Lockheed Martin Corporation Endowed Professorship Fund, in part by the NASA award, and generous gifts from NVIDIA and Google.

%\begingroup
%\tiny
%\renewcommand{\baselinestretch}{0.9}

%\balance

%\IEEEtriggeratref{10}
\bibliographystyle{IEEEtran}
\bibliography{References,Hassan}
%\endgroup

\begin{IEEEbiography}[{\includegraphics[width=1.09in,height=1.2in]{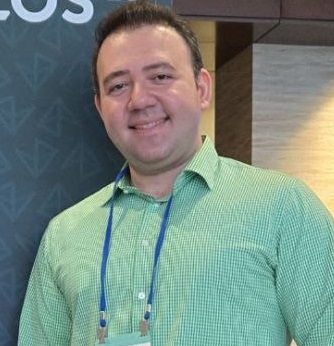}}]{Farzad Razi} (M'25) received the B.Sc. degree in Computer Engineering (Hardware Engineering) from the University of Isfahan, Isfahan, Iran, in 2013, and the M.Sc. degree in Computer Engineering (Computer Architecture) from Shahid Beheshti University, Tehran, Iran, in 2016, graduating as a top-ranking student in both programs. He received the Ph.D. degree in Computer Engineering (Computer Systems Architecture) from the University of Tehran, Tehran, Iran, in 2023, where he ranked first among his Ph.D. cohort.
Dr. Razi received the Best Paper Award as first author at IEEE/ACM ISLPED'26.
He is currently a Postdoctoral Associate Researcher in the Department of Electrical and Computer Engineering, University of Minnesota, Minneapolis, MN, USA. He works on emerging computing, including in-memory computing, stochastic computing, low-power and high-performance design, and fault-tolerant architectures for AI and edge computing.
\end{IEEEbiography}

\begin{IEEEbiography}[{\includegraphics[width=1.3in,height=1.25in,clip,keepaspectratio]{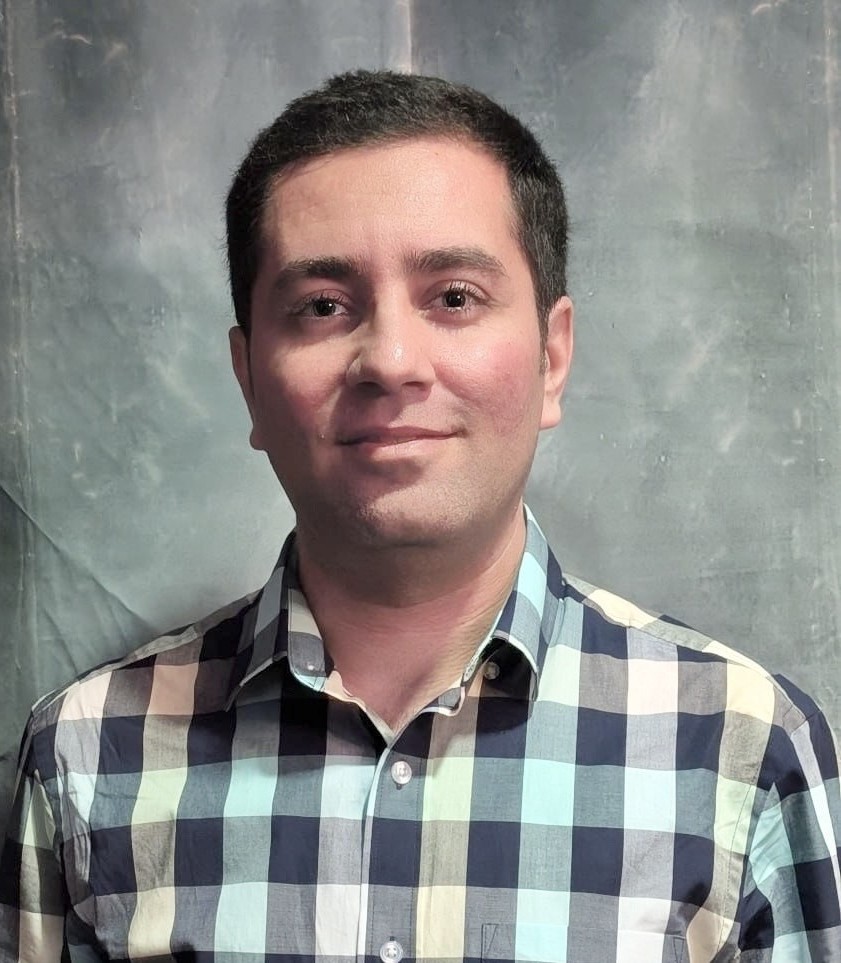}}]{Mehran Moghadam} (S’22) received the B.Sc. degree in Computer Engineering, and M.Sc. degree in \textit{Computer Systems Architecture} from the University of Isfahan, Iran, in 2010 and 2016. He graduated as one of the top-ranking students in both programs.
In 2022, he began his Ph.D. studies in Computer Engineering at the School of Computing and Informatics, University of Louisiana at Lafayette, Lafayette, LA, USA. In 2024, he transferred to the Department of Electrical, Computer, and Systems Engineering at Case Western Reserve University, Cleveland, OH, USA, to continue pursuing his Ph.D. in Computer Engineering.
He became a finalist in the ACM SIGBED Student Research Competition (SRC) at ESWEEK and ICCAD in 2024 and was selected as a Young Fellow in DAC 2024 and 2026. %He has authored/co-authored more than 30 peer-reviewed papers in top EDA venues.
Mehran received the Best Paper Award at ISLPED'26.
His research interests include emerging and unconventional computing paradigms, such as energy-efficient stochastic computing models, real-time and highly-accurate hyperdimensional computing systems, bit-stream processing, robust in-memory arithmetic computation, and low-power in/near-sensor computing designs for edge AI and next-generation wearable platforms. 
\end{IEEEbiography}

\begin{IEEEbiography}[{\includegraphics[width=1.09in,height=1.2in]{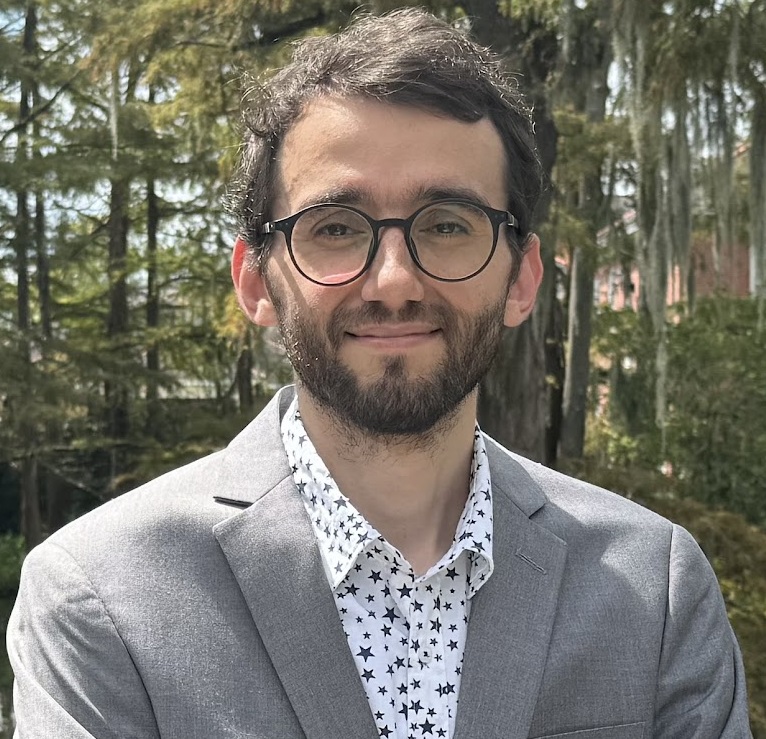}}]{Sercan Aygun} (S’09-M’22-SM'24) received a B.Sc. degree in Electrical \& Electronics Engineering and a double major in Computer Engineering from Eskisehir Osmangazi University, Turkey, in 2013. He completed his M.Sc. degree in Electronics Engineering from Istanbul Technical University in 2015 and a second M.Sc. degree in Computer Engineering from Anadolu University in 2016. Dr. Aygun received his Ph.D. in Electronics Engineering from Istanbul Technical University in 2022. %Dr. Aygun’s Ph.D. work has appeared in several Ph.D. Forums of top-tier conferences, such as DAC, DATE, ASP-DAC, and ESWEEK.
Dr. Aygun received the Best Scientific Research Award of the ACM SIGBED Student Research Competition (SRC) ESWEEK 2022, the Best Paper Award at GLSVLSI'23 \& ISLPED'26, and the Best Poster Awards at GLSVLSI'24 \& DCAS'26. Dr. Aygun's Ph.D. work was recognized with the Best Scientific Application Ph.D. Award by the Turkish Electronic Manufacturers Association and was also ranked first nationwide in the Science and Engineering Ph.D. Thesis Awards by the Turkish Academy of Sciences. He is currently an Assistant Professor at the School of Computing and Informatics, University of Louisiana at Lafayette, Lafayette, LA, USA. He works on tiny machine learning and emerging computing, including stochastic and hyperdimensional computing. \end{IEEEbiography}

\begin{IEEEbiography}[{\includegraphics[width=1.3in,height=1.25in,clip,keepaspectratio]{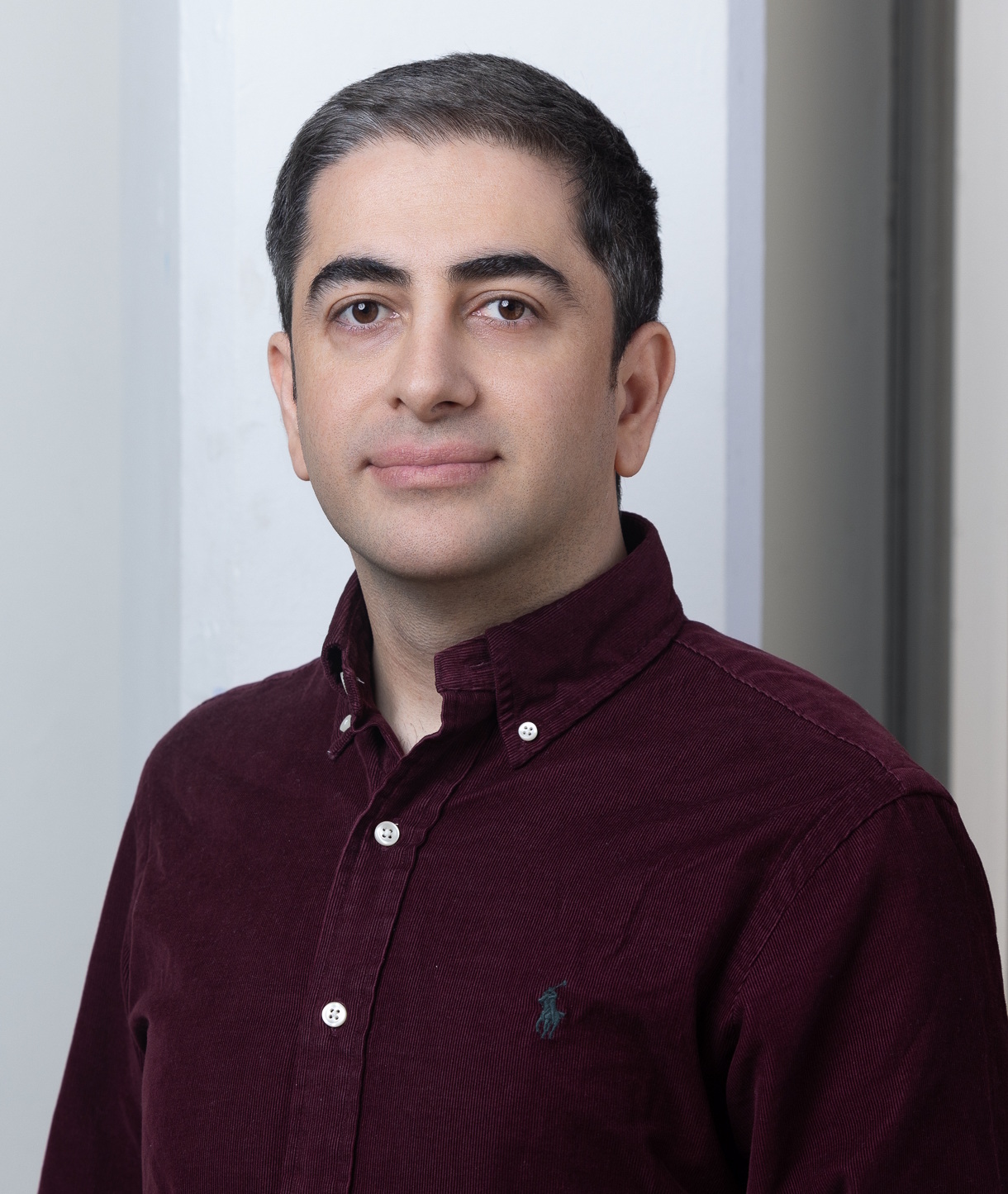}}]{M. Hassan Najafi} (S’15-M’18-SM'23) received the B.Sc. degree in Computer Engineering from the University of Isfahan, Iran, the M.Sc. degree in Computer Architecture from the University of Tehran, Iran, and the Ph.D. degree in Electrical Engineering from University of Minnesota, Twin Cities, USA, in 2011, 2014, and 2018, respectively. He was an Assistant Professor at the School of Computing and Informatics, University of Louisiana at Lafayette, Lafayette, LA, USA, from 2018 to 2024. He is currently an Associate Professor at the Electrical, Computer, and Systems Engineering Department at Case Western Reserve University, Cleveland, OH, USA. His research interests include stochastic and approximate computing, unary processing, in-memory computing, and hyperdimensional computing. He has authored/co-authored more than 120 peer-reviewed papers and has been granted 12 U.S. patents, with more pending. Dr. Najafi received the NSF CAREER Award in 2024, the Best Paper Award at ICCD’17, GLSVLSI'23, and ISLPED'26, the Best Poster Awards at GLSVLSI'24 and DCAS'26, the 2018 EDAA Outstanding Dissertation Award, and the Doctoral Dissertation Fellowship from the University of Minnesota. He has been an editor for the IEEE Journal on Emerging and Selected Topics in Circuits and Systems and a Technical Program Committee Member for many EDA conferences. Dr.~Najafi is a Senior Member of IEEE and a Senior Member of the U.S. National Academy of Inventors (NAI). \end{IEEEbiography}

\begin{IEEEbiography}[{\includegraphics[width=1.0in,height=1.2in]{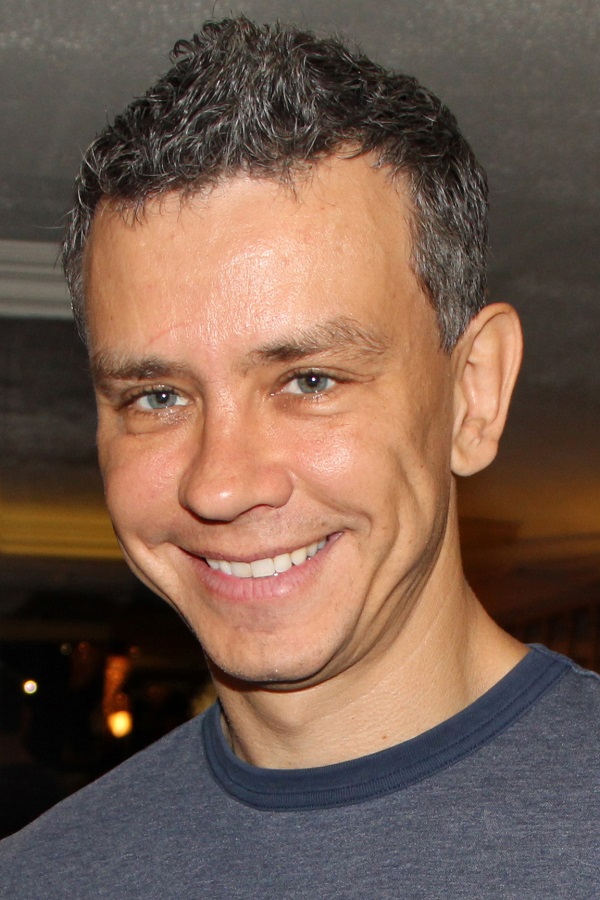}}]{Marc D. Riedel} (SM'12) received the B.Eng. degree
in electrical engineering from McGill University,
Montreal, QC, Canada, and the M.Sc. and Ph.D.
degrees in electrical engineering from the California
Institute of Technology (Caltech), Pasadena, CA,
USA.
From 2004 to 2005, he was a Lecturer of computation and neural systems at Caltech. He was at Marconi Canada, CAE Electronics, Toshiba, and
Fujitsu Research Labs. He is currently a Full Professor in the Department of Electrical and Computer Engineering, University of Minnesota, Minneapolis, MN, USA, where he is a
member of the Graduate Faculty of biomedical informatics and computational
biology.
Dr. Riedel was a recipient of the Charl H. Wilts Prize for the Best Doctoral
Research in Electrical Engineering at Caltech, the Best Paper Award at the
Design Automation Conference, and the U.S. National Science Foundation
CAREER Award and has been named an Oracle Research Fellow. 
His research interests spans emerging computing, stochastic computing, synthetic biology, and molecular and DNA-based computing.
%received the B.Eng. degree from McGill University, Montreal, QC, Canada, in 1995, and the M.S. and Ph.D. degrees in Electrical Engineering from the California Institute of Technology, Pasadena, CA, USA, in 1997 and 2004, respectively. He is currently a Full Professor in the Department of Electrical and Computer Engineering, University of Minnesota, Minneapolis, MN, USA, which he joined as an Assistant Professor in 2006. He is a recipient of the National Science Foundation CAREER Award and has been named an Oracle Research Fellow. His research spans emerging computing, stochastic computing, synthetic biology, and molecular and DNA-based computing.
\end{IEEEbiography}

\end{document}